\documentclass[useAMS,usenatbib]{mn2e}
\usepackage{graphicx,amssym}
\newcommand{\bc}{\begin{center}}
\newcommand{\ec}{\end{center}}

\title[Modelling galaxy clustering]
      {Modelling galaxy clustering in a high resolution simulation of structure
       formation}
\author[L.Wang, C.Li, G.Kauffmann, G.~De Lucia]
       {Lan Wang$^{1,}$$^2$$\thanks{Email: wanglan@mpa-garching.mpg.de}$,
	Cheng Li$^{2,3,4}$,
        Guinevere Kauffmann$^2$,
	Gabriella De Lucia$^2$   
        \\      
	$^1$Department of Astronomy, Peking University, Beijing 100871, China\\
        $^2$Max--Planck--Institut f\"ur Astrophysik, 
        Karl--Schwarzschild--Str. 1, D-85748 Garching, Germany\\
      	$^3$Center for Astrophysics, University of Science
      and Technology of China, Hefei, Anhui 230026, China\\
	$^4$The Partner Group of MPI f\"ur Astrophysik,
      Shanghai Astronomical Observatory,
      Nandan Road 80, Shanghai 200030, China}
\begin{document}

\date{Accepted 2006 ???? ??. 
      Received 2006 ???? ??; 
      in original form 2006 ???? ??}

\pagerange{\pageref{firstpage}--\pageref{lastpage}} 
\pubyear{2005}

\maketitle

\label{firstpage}

\begin{abstract}
We use  the {\it Millennium Simulation}, a 10 billion particle  
simulation of the growth of cosmic structure,
to construct a new  model of galaxy clustering. We adopt a  methodology that 
falls midway between the traditional  semi-analytic approach and the
halo occupation distribution (HOD) approach. In our model,  we  
adopt the positions and velocities of the galaxies that are predicted by
following the orbits and merging histories of the substructures
in the simulation. Rather than using star formation and
feedback `recipes' to specify  the  physical properties of the
galaxies, we adopt parametrized functions to relate these properties 
to the quantity 
$M_{infall}$, defined as the mass of the halo 
at the epoch when the galaxy was last the
central dominant object in its own halo. We test whether these parametrized
relations allow us to recover the  basic statistical properties of
galaxies in the semi-analytic catalogues, including the luminosity function,
the stellar mass function and the shape and
amplitude of the  two-point correlation function evaluated in different
stellar mass and luminosity ranges. We then use our model to interpret
recent measurements of these quantities from  Sloan
Digital Sky Survey data.
We derive relations between the  luminosities and the stellar
masses of galaxies in the local Universe  and 
their host halo masses. Our results are in excellent agreement
with recent determinations  of these relations by Mandelbaum et al    
using galaxy-galaxy weak lensing measurements from the SDSS.

\end{abstract}

\begin{keywords}
   galaxies: fundamental parameters -- galaxies: haloes -- galaxies: 
	distances and redshifts -- cosmology: theory -- 
	cosmology: dark matter -- cosmology: large-scale structure
\end{keywords}

%%%%%%%%%%%%%%%%%%%%%%%%%%%%%%%%%%%%%%%%%%%%%%%%%%%%%%%%%%%%%%%%%%%%%%%%%%%%%%
\section{Introduction}
\label{sec:intro}

According to the current standard paradigm, galaxies form and reside inside 
extended dark matter haloes. Three different approaches have been used to 
model the link between the properties of galaxies and the dark matter 
haloes in which they are found. One approach is to carry out 
N-body + hydrodynamical simulations
that include both gas and dark matter\citep{katz1996, pearce2001, white2001,
yoshikawa2001}. Another approach is to combine N-body
simulations with  simple prescriptions,
taken directly from semi-analytic models of galaxy 
formation\citep{kauffmann1999}, to track gas cooling and 
star formation in galaxies. The third method is the so called Halo Occupation 
Distribution (HOD) approach, which aims to provide a purely statistical 
description of how dark matter haloes are populated  by galaxies.

Typical HOD models are constructed by specifying the number of galaxies $N$
that populate a dark matter halo of mass $M$ as well as the distribution
of galaxies within these haloes\citep{kauffmann1997,peacock2000,seljak2000,benson2000,berlind2002,berlind2003}. More recent 
models have concentrated on the so-called conditional 
luminosity function $\Phi(L|M) dL$, which gives the number of galaxies of 
luminosity $L$ that reside in a halo of mass $M$\citep{yang2003}. 
Most HOD models also distinguish between ``central'' galaxies, located at the 
centres of dark matter haloes
and ``satellite'' galaxies, which are usually assumed to have the same density 
profile as the dark matter within the halo. Physically, this is supposed
to reflect the fact that gas cools and accumulates at the halo centres until
the halo merges with a larger structure. With this approach, the models 
can be used to explore the parameters that are required to match 
simultaneously the galaxy luminosity function as well as the the luminosity, 
colour and morphology dependences of the correlation 
function\citep{bosch2003,zehavi2005, yang2005}. Other papers have used HOD 
models to explore the detailed shape of two-point correlation 
function\citep{zehavi2004} as well as  higher order correlation 
functions\citep{wang2004}. 

N-body simulations can now be carried out with high enough resolution to 
track the histories of individual substructures (subhaloes) within the 
surrounding halo\citep{springel2001}. It is thus becomes possible to specify  
the positions and velocities of galaxies within a halo in
a dynamically consistent way, rather 
than assuming a profile or form for the velocity distribution. Galaxy 
clustering statistics that are computed using the full information available 
from these high resolution simulations should in principle  be more accurate
and robust. 

Caution must be exercised, however, when only using subhaloes as tracers of
galaxies in high resolution simulations, as has been recently done by
\citet{vale2004,vale2005,conroy2005}.
In standard models of galaxy formation, when a galaxy is accreted by a larger
system such as a cluster, its surrounding gas is shock--heated to high 
temperatures. Star formation then terminates as the internal
gas supply of the galaxy is used up. The stellar masses of satellite galaxies 
only change by a small amount after they are accreted,
while their luminosities dim due to aging of their stars. In contrast, 
the dark matter haloes
surrounding the satellites gradually lose mass as their outer regions are
tidally stripped \citep{delucia2004a}. Near the centres of the halos, 
most of the substructures have been completely destroyed. \citet{gao2004}
have shown that the radial distribution of subhaloes is much less
centrally concentrated than the radial distribution of galaxies
predicted by simulations that follow the full orbital  and 
merging histories of these systems. {\footnote {Note that the simulations
analyzed by \cite{conroy2005} are significantly higher resolution
than the ones analyzed in this paper, but are much smaller in volume.
As discussed in their paper, the problem of disrupted subhaloes
is not likely to be a significant problem for galaxies in the range
of luminosities considered in their analysis.}}

In this paper, we make use of the {\it Millennium Simulation}, a 10 billion
particle simulation of the growth of cosmic structure,
to construct a new  model of galaxy clustering. We adopt a  methodology that 
falls in  between the semi-analytic approach, which tracks galaxy formation 
`ab initio' within the simulation, and the HOD approach, which only provides
a statistical description of how galaxies are related to the
underlying dark matter density distribution. In our approach, we
adopt the positions and velocities of the galaxies as predicted by
following the orbits and merging histories
of the substructures in the simulation. Rather than using star formation and
feedback `recipes' to calculate how the  physical properties of the
galaxies such as their luminosities or stellar masses evolve with time, we 
adopt parametrized functions to relate these properties to the quantity 
$M_{infall}$, defined as the mass of the halo 
at the epoch when the galaxy was last the
central dominant object. For central galaxies at the present
day, $M_{infall}$ is simply the present day halo mass, but
for satellite galaxies, it is the mass of the halo when the
galaxy was first accreted by a larger structure. 

This approach has the advantage of the semi-analytic models in that it provides
 very accurate positions and velocities for all the galaxies
in the simulation. It maintains the simplicity of the HOD
approach, because it bypasses the need to incorporate detailed treatment
of star formation and feedback processes.
Our aim in developing these models is to use them as a means
of constraining the relation between galaxy physical properties 
and halo mass directly from observational data, not as a means
of understanding the physics of galaxy formation.

The paper is organized as follow: we first introduce the Millennium Run and 
the methodology used for identifying haloes, subhaloes and galaxies in this 
simulation. We then study the relation between luminosity/stellar mass and  
$M_{infall}$ in mock galaxy catalogues constructed using these simulations. 
In Sec.~\ref{sec:para} we introduce a parametrization for these relations
and show that we are able to recover basic statistical quantities such as the
galaxy luminosity/mass function and the shape and amplitude of the two-point
correlation function in different luminosity/mass bins. We also investigate the
the effect of changing the parameters of the relation on the luminosity
function and correlation function. In Sec.~\ref{sec:sdss} we apply the method 
to real data on the clustering of galaxies as a function of luminosity and
stellar mass\citep{li2005a} derived from the Sloan Digital Sky Survey (SDSS).
Finally, we discuss our results and present our conclusions.

% %%%%%%%%%%%%%%%%%%%%%%%%%%%%%%%%%%%%%%%%%%%%%%%%%%%%%%%%%%%%%%%%%%%%%%%%%%%%%
\begin{figure*}
\bc
\hspace{-1.6cm}
\resizebox{17.cm}{!}{\includegraphics{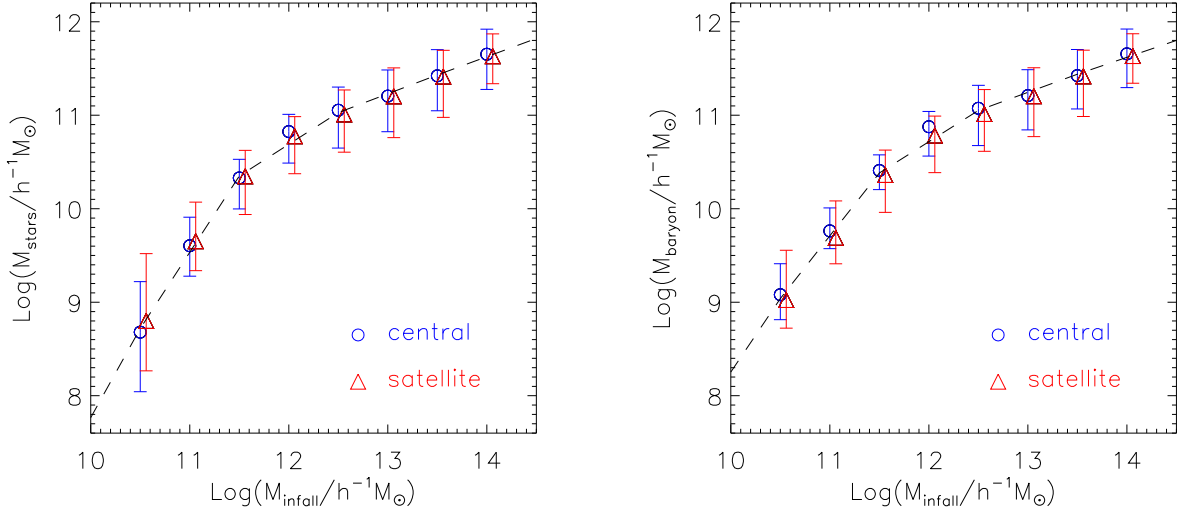}}\\%
\caption{ 
 Relations between stellar mass, baryonic mass and $M_{infall}$ calculated 
from the semi-analytic galaxy catalogues. 
Open circles represent central galaxies and triangles are for
satellite galaxies. Error bars indicate  the $95$ percentiles of the mass
distribution at the given value of $M_{infall}$.
Dashed lines show the double power law parametrized fit 
to the median value of relations
for the galaxy sample as whole. }
\label{fig:MMinfall}
\ec
\end{figure*}

\section{The Simulation}
\label{sec:simulation}

The \emph{Millennium Simulation}\citep{springel2005} used in this study, is 
the largest simulation of cosmic structure growth carried out so far. The 
cosmological parameters values in the simulation are consistent with recent 
determinations from a combined analysis of the 2dFGRS\citep{colless01} and 
first year WMAP data \citep{spergel03}. A flat $\Lambda$CDM cosmology is 
assumed with $\Omega_{\rm m}=0.25$, $\Omega_{\rm b}=0.045$, $h=0.73$, 
$\Omega_\Lambda=0.75$, $n=1$, and $\sigma_8=0.9$. The simulation follows 
$N= 2160^3$ particles of mass $8.6\times10^{8}\,h^{-1}{\rm M}_{\odot}$ from 
redshift $z=127$ to the present day, within a comoving box of 
$500\, h^{-1}$Mpc on a side.  

Full particle data are stored at 64 output times. For each output, haloes
are identified using a friends-of-friends (FOF) group-finder. Substructures
(or subhaloes) within a FOF halo are located using the   
SUBFIND algorithm of \citet{springel2001}. After finding all haloes and 
subhaloes at all output snapshots, merging trees are built describing in 
detail how these systems merge and grow as the universe 
evolves. Since structures merge hierarchically in CDM universes, for any given 
halo, there can be several progenitors, but in general each
halo or subhalo only has one descendant.
Merger trees are thus constructed by defining a unique descendant 
for each halo and subhalo. Through those merging trees, we are able to 
follow the history of haloes/subhaloes, as well as the galaxies inside them.

Once a halo appears in the simulation, it is assumed that a galaxy 
begins to form within it. As the simulation evolves, the halo may merge with 
a larger structure and become a subhalo, while the galaxy becomes a satellite 
galaxy. The galaxy's position and velocity are specified by the position and 
velocity of the most bound particle of its host halo/subhalo.  
Even if the subhalo hosting the galaxy is tidally disrupted, the position
and velocity of the galaxy is still traced through this most bound particle.
We will refer to these galaxies without subhaloes as ``orphaned'' systems. 
Galaxies thus
only disappear from the simulation if they merge with another galaxy.
The time taken for an orphaned galaxy to merge with the central object
is given by the time taken for  dynamical friction to erode its orbit, 
causing it to spiral into the centre and merge.
The satellite orbits are thus tracked directly until the subhalo is disrupted;
thereafter, the time taken for the galaxy to reach the centre is calculated
using the standard Chandrasekhar formula.
 
In this paper, we will parameterize quantities such as galaxy luminosity
and stellar mass as a function of the quantity $M_{infall}$, which is defined
as the virial mass of the halo hosting the galaxy at the epoch when
it was last the central galaxy of its own halo.  The Millennium simulation
catalogues include haloes down to a resolution limit of 20 particles, 
which yields a minimum halo mass of $2 \times 10^{10}h^{-1}M\odot$.
In our study, we only consider galaxies with $M_{infall}$ greater than  
$10^{10.5}h^{-1}M\odot$.(Note that $M_{infall}$ is simply the
virial mass of the host halo for central galaxies at the present day.)
This results in a total sample of $11761178$ galaxies within the 
simulation volume.

% %%%%%%%%%%%%%%%%%%%%%%%%%%%%%%%%%%%%%%%%%%%%%%%%%%%%%%%%%%%%%%%%%%%%%%%%%%%%%
\section{The relations between M$_{infall}$, stellar mass
and luminosity in the semi-analytic galaxy catalogues}
\label{sec:relation}

In the following two sections we use the semi-analytic galaxy catalogues 
constructed from the Millennium simulation by
\citep{croton2005} (http://www.mpa-garching.mpg.de/galform/agnpaper/)
to study how galaxy properties such as  stellar
mass, baryonic mass (i.e. stellar mass+ cold gas mass) 
and luminosity depend on $M_{infall}$, the mass of the halo
in which the galaxy was last a central object. We construct  
parametrized relations between these quantities and $M_{infall}$ 
that match the relations found in the mock catalogue. 
We then show that our parametrization allows us to recover both the
luminosity/mass functions of the simulated galaxies
and the shape, amplitude and mass/luminosity dependence of the
two-point correlation functions. \cite{croton2005} have shown that 
their catalogues provide a good match to the observed galaxy luminosity 
function and the clustering properties of galaxies,
so we believe that it is a reasonable to use these catalogues as a way
of motivating and testing our simple parametrizations.  

In Fig.~\ref{fig:MMinfall} we plot the relations between $M_{infall}$ and  
galaxy stellar mass ($M_{stars}$) and baryonic mass ($M_{baryon}$).  
We show results for present-day central
galaxies in blue and satellite galaxies in red.
Error bars indicate the 95th percentiles of the distributions.
As we will show, the relations between $M_{infall}$ and
$M_{stars}$/ $M_{baryon}$ are well described by a double power law.
The crossover point between the two power laws is at a halo mass of  
$\sim3\times10^{11}h^{-1}M_{\odot}$, which corresponds to a galaxy 
with stellar mass of  around $10^{10}h^{-1}M_{\odot}$. In less massive  
haloes, supernova feedback acts to prevent gas from cooling and forming stars 
as efficiently as in high mass haloes. In massive haloes, the cooling times 
become longer and a smaller fraction of the baryons are predicted to cool and 
form stars. In addition, in the models
of \cite{croton2005},  heating from AGN also acts to suppress
cooling onto high mass galaxies.

\begin{figure}
\bc
\hspace{-0.6cm}
\resizebox{8.5cm}{!}{\includegraphics{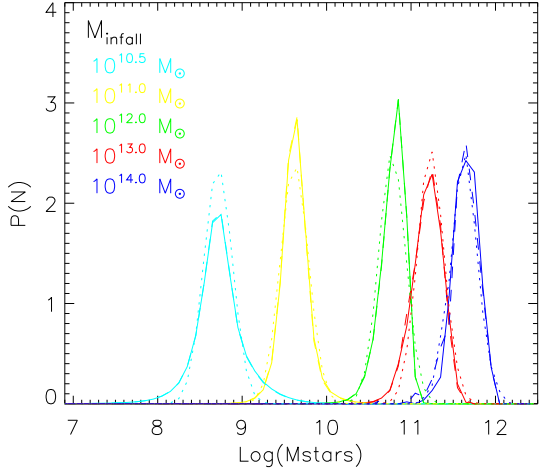}}\\%
\caption{The distribution in stellar mass for different $M_{infall}$ bins 
increasing from $10^{10.5}h^{-1}M\odot$(left) to $10^{14.0}h^{-1}M\odot$(right). 
Solid and dashed lines are for central and satellite galaxies (note that they 
lie on  top of each other for three lower mass bins).
Dotted lines indicate the Gaussian fits to the distributions that are used 
in our parametrized model.}
\label{fig:scatter}
\ec
\end{figure}

Fig.~\ref{fig:scatter} shows that the distribution of $M_{stars}$
at a given value of $M_{infall}$ is well-described by a log-normal
function. The width of the lognormal depends weakly on halo mass with 
a maximum dispersion  $\sigma$ of 0.2 dex 
at $ M_{infall} = 10^{10.5}h^{-1}M_{\odot}$ and a minimum $\sigma$  
of 0.1 dex  at $M_{infall}= 10^{11.5}h^{-1}M_{\odot}$.  
The relations depend very little on whether the galaxy is
a central or satellite system. The dispersion around the relations is also
similar for the two types(in Fig.~\ref{fig:scatter}, the solid 
and dashed lines for central and satellite galaxies lie almost on top
of each other).
The similarity in the $M_{infall}$- $M_{stars}$ relations 
between satellite and centrals may be regarded as something of
a coincidence.
Although  there is little change in the
stellar/baryonic component of the galaxy after it
falls into a larger structure, halos of the same mass at different times
have different circular velocities and hence different cooling and 
and star formation efficiencies. As we will show later, we obtain better fits
to the observational data if we allow the relations  between
central and satellites galaxies to differ.

\begin{figure}
\bc
\hspace{-0.6cm}
\resizebox{8.5cm}{!}{\includegraphics{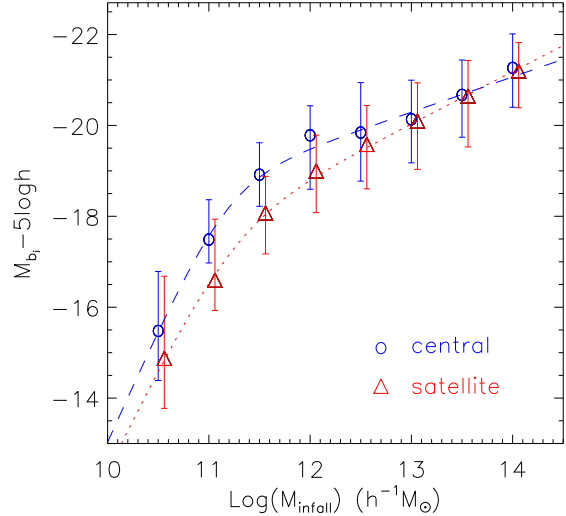}}\\%
\caption{The same as in Fig.~\ref{fig:MMinfall},  but  
luminosity (represented by magnitude of $b_j$ band) is
plotted as a function of  $M_{infall}$. 
Dashed(dotted) lines show the double power law fits to the relation  
for central(satellite) galaxies.}
\label{fig:L}
\ec
\end{figure}

Fig.~\ref{fig:L} shows the relation between luminosity and $M_{infall}$.
It also can be fit by a double power law, but the difference between central
and satellite galaxies is much more obvious. At a given value of $M_{infall}$
central galaxies are more luminous than satellites because they are
forming stars at higher rates and their stellar mass-to-light ratios are lower.
The difference between central and satellite
galaxies becomes very small at large values of $M_{infall}$.
This can be understood as a simple consequence
of hierarchical structure formation:  massive haloes were formed
more recently than less massive haloes and subhaloes with large masses
are likely to have been accreted relatively recently.
Massive  satellite galaxies have therefore not been  
satellites for long and thus have mass-to-light ratios that are
more similar to their central counterparts. In addition the
Croton et al models include a ``radio AGN  mode'' of feedback,
which acts to suppress cooling onto the most massive galaxies.
This also acts to reduce the difference between central
and satellite galaxy colours and mass-to-light ratios.

% %%%%%%%%%%%%%%%%%%%%%%%%%%%%%%%%%%%%%%%%%%%%%%%%%%%%%%%%%%%%%%%%%%%%%%%%%%%%%
\section{Parametrizations and tests}
\label{sec:para}

 \begin{table*}
 \caption{Best-fit parameter values for the relations between
$M_{infall}$ and $M_{stars}$, $M_{baryon}$ and $L$ as derived
from the semi-analytic galaxy catalogues of \citet{croton2005}.}
\begin{center}
 \begin{tabular}{cccccccc} \hline
   &         &  $M_0(h^{-1}M_{\odot})$ & $\alpha$ & $\beta$ & $log(k)$  & $\sigma$& $\tilde{\chi}^2$\\ \hline
 $M_{stars}$ & total & 3.16$\times10^{11}$ & 0.39 & 1.92 & 10.35  & 0.156& 0.0146\\
             & central & 3.16$\times10^{11}$ & 0.39 & 1.96 & 10.35  & 0.148& 0.0240\\
             & satellite & 3.16$\times10^{11}$ & 0.39 & 1.83 & 10.34  & 0.167& 0.0057\\ \hline
$M_{baryon}$ & total & 3.61$\times10^{11}$ & 0.36 & 1.59 & 10.44  & 0.147& 0.0415\\
             & central & 3.61$\times10^{11}$ & 0.35 & 1.59 & 10.46  & 0.133& 0.0542\\
             & satellite & 3.61$\times10^{11}$ & 0.37 & 1.59 & 10.40  & 0.162& 0.0273\\ \hline
$L(M_{bj})$          & total & 1.49$\times10^{11}$ & 0.36 & 1.90 & 7.14  & 0.215& 0.0360\\
             & central & 1.49$\times10^{11}$ & 0.31 & 1.99 & 7.25  & 0.169& 0.1250\\
             & satellite & 1.49$\times10^{11}$ & 0.46 & 1.81 & 6.90  & 0.189& 0.0359\\ \hline
 \end{tabular}
\end{center}
 \end{table*}

\subsection {Functional form}

We use a two--power--law model of the following form
to fit the median value of the relations 
between $M_{stars}$, $M_{baryon}$, $L$ and $M_{infall}$: 
\begin{displaymath}
{x} = \frac{2}{(\frac{{M}_{infall}}{{{M}_{0}}})^{-\alpha}+(\frac{{ M}_{infall}}{{{M}_{0}}})^{-\beta}}{\times}{k},
\end{displaymath}
where $x$ denotes $M_{stars}$,$M_{baryon}$ or $L$, and the relation between 
luminosity $L$ and  $b_j$ band magnitude is given by:
\begin{displaymath}
{{M}_{b_{j}}-{5}{\log}{h}}={-2.5}{\log}{L}
\end{displaymath}
We fit these relations for central and satellite galaxies separately,
as well as for the galaxy population as a whole. We will later test
whether separate fits to the central galaxies and satellites make  
significant difference to our results. We also assume that the dispersion 
around the median value has a lognormal form. 

Table 1 lists the parameters of the best--fitting models for
the relations between $M_{infall}$ and $M_{stars}$, $M_{baryon}$
and $L$. The models have five parameters. 
\begin {enumerate}
\item
$M_{0}$ is the critical mass/luminosity at which the slope of  the relation 
changes. When we fit satellite and central galaxies separately, we find
almost exactly the same values for this parameter (even for
luminosity, the difference is less than 20\%).    
We therefore fix $M_0$ at the best-fit value
for the galaxy sample as a whole.
\item
 $\alpha$ and $\beta$ describe the slope of the relations  at 
high and low values of $M_{infall}$. 
\item
$k$ is a normalization constant.            
\item
We have calculated the interval in 
$\log M_{stars}$, $\log M_{baryon}$
and $\log L$ that encloses the central 68\% of the probability distribution
for  8 different values of $\log M_{infall}$
from $10^{10.5}h^{-1}M\odot$ to $10^{14}h^{-1}M\odot$, with step 0.5 dex.
We then calculate the average of these 8 values and the value
of $\sigma$ quoted in Table 1 is 0.5 times this number.
\end {enumerate}

The resulting model fits are plotted as dashed and dotted lines in   
Fig.~\ref{fig:MMinfall} and Fig.~\ref{fig:L}.  
The quality of the fit is given in the last column of Table 1 and is   
calculated as:
\begin{displaymath}
{\tilde{\chi}^{2}} = \sum{(\frac{{x_{fit}}-{x_{SAM}}}{{x_{SAM}}})^{2}}
\end{displaymath}
where $x$ represents $M_{stars}$, $M_{baryon}$, $L$ for each relation, and 
the sum is over the $8$ mass bins with $10^{10.5}h^{-1}M_{\odot}{\leq} M_{infall}\leq10^{14.0}h^{-1}M_{\odot}$. 

\begin{figure*}
\bc
\hspace{-1.4cm}
\resizebox{17cm}{!}{\includegraphics{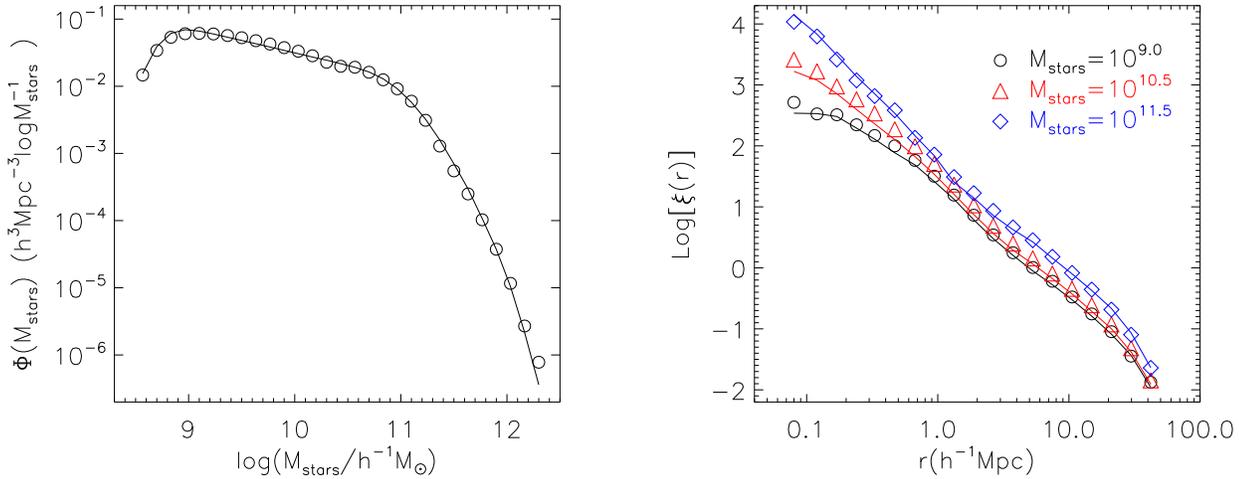}}\\%
\caption{ 
Results from the comparison of the parametrized model and the
semi-analytic galaxy catalogue. The left panel shows the stellar
mass function. The right panel shows correlation functions for three 
different stellar mass bins: $10^{9}h^{-1}M\odot$, $10^{10.5}h^{-1}M\odot$ and 
$10^{11.5}h^{-1}M\odot$. Symbols are for the semi--analytic galaxy catalogue.
Solid lines are for our parametrized models.} 
\label{fig:SMFcorr}
\ec
\end{figure*}

\begin{figure*}
\bc
\hspace{-1.4cm}
\resizebox{17cm}{!}{\includegraphics{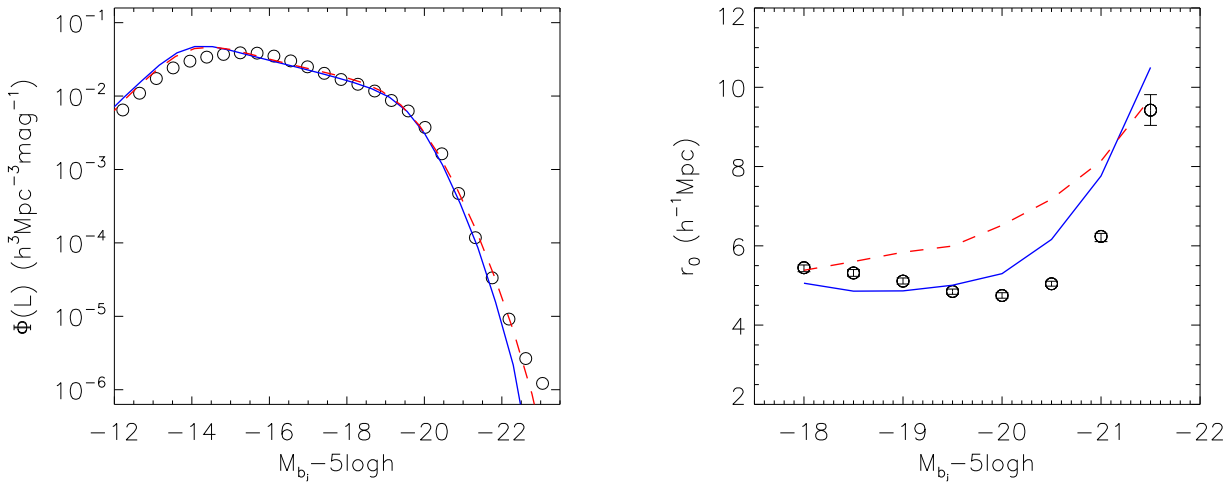}}\\%
\caption{
Same as Fig. 4, except for the luminosity function (left) and the   
correlation function as a function of absolute magnitude (right).
Symbols are the semi-analytic results, error bars in the right panel
are the boot strap error of correlation length for the semi-analytic model. 
Solid lines are for parametrized  models where relations for central galaxies  
and satellite galaxies are fit separately. The dashed lines are for models
where the fit is for the galaxy population as a whole.}

\label{fig:LFcorr}
\ec
\end{figure*}

\subsection {Tests}
The next step is to see whether these parametrized relations 
allow us to recover the basic statistical properties of the simulated galaxy
catalogue, such as the mass/luminosity function and the mass/luminosity
dependence of the two point correlation function. When fitting to the 
quantities $M_{star}$ and $M_{baryon}$, we do not distinguish between
central and satellite galaxies because the relations are almost the same 
for both. When fitting to galaxy luminosity,
we do allow $\alpha$, $\beta$, $k$ and $\sigma$ to vary between
central and satellite galaxies, but $M_0$ remains fixed for both. 
Note that the positions and the velocities of the galaxies are exactly
the same as specified in the semi-analytic galaxy catalogues; the
parametrized relations between galaxy mass/luminosity
and $M_{infall}$  simply provide us with a alternative way of
specifying the {\em properties} of the galaxies.

Fig.~\ref{fig:SMFcorr} and Fig.~\ref{fig:LFcorr} show the results of our test.
Symbols show results calculated directly from the semi-analytic 
galaxy catalogues and lines are from our parametrized model. The stellar mass 
function is well reproduced, and we can also recover 
the correlation for different stellar mass bins: $10^{9}h^{-1}M_{\odot}$, 
$10^{10.5}h^{-1}M_{\odot}$ and $10^{11.5}h^{-1}M_{\odot}$(Fig.~\ref{fig:SMFcorr}). 
For luminosity, the parametrized model is not quite as successful.
Although the luminosity function is well-reproduced, there are some
discrepancies in the dependence of the clustering amplitude on
luminosity (solid-blue curve in Fig.~\ref{fig:LFcorr}).  
Part of the reason for this discrepancy is that our parametrization of 
the $L-M_{infall}$ relation has somewhat larger $\chi^2$ than
the $M_{stars}-M_{infall}$ relation (see Table.1). 
In addition, in order to reproduce the clustering trends as
a function of luminosity, it is critical to fit the relation
for central and for satellites galaxies separately. If we apply a single
relation for both kinds of galaxies, we obtain the red-dashed
curve in Fig.~\ref{fig:LFcorr}, which is even more discrepant.
Our results suggest that in order to reproduce the clustering
dependence on luminosity in a more exact way, one would need to introduce 
an additional dependence of the $L-M_{infall}$ relation on the
parameter $t_{infall}$, the {\em time}  when the galaxy was 
last the central object of its own halo. This does not appear to
be necessary in order to reproduce the stellar mass dependence of galaxy
clustering. The reason for this difference is because
the optical light from galaxies, unlike
their stellar mass, is heavily influenced  by the contribution from
the youngest stars, which have lifetimes which are short compared to
the age of the Universe. Once a galaxy becomes a satellite,
it will fade in luminosity even though its stellar mass remains
approximately constant. For the sake of simplicity,
we will not introduce $t_{infall}$ as an additional parameter
in this paper, but we will come back to this in future work in which
we consider the colour-dependence of galaxy clustering.

\begin{figure*}
\bc
\hspace{-1.4cm}
\resizebox{17cm}{!}{\includegraphics{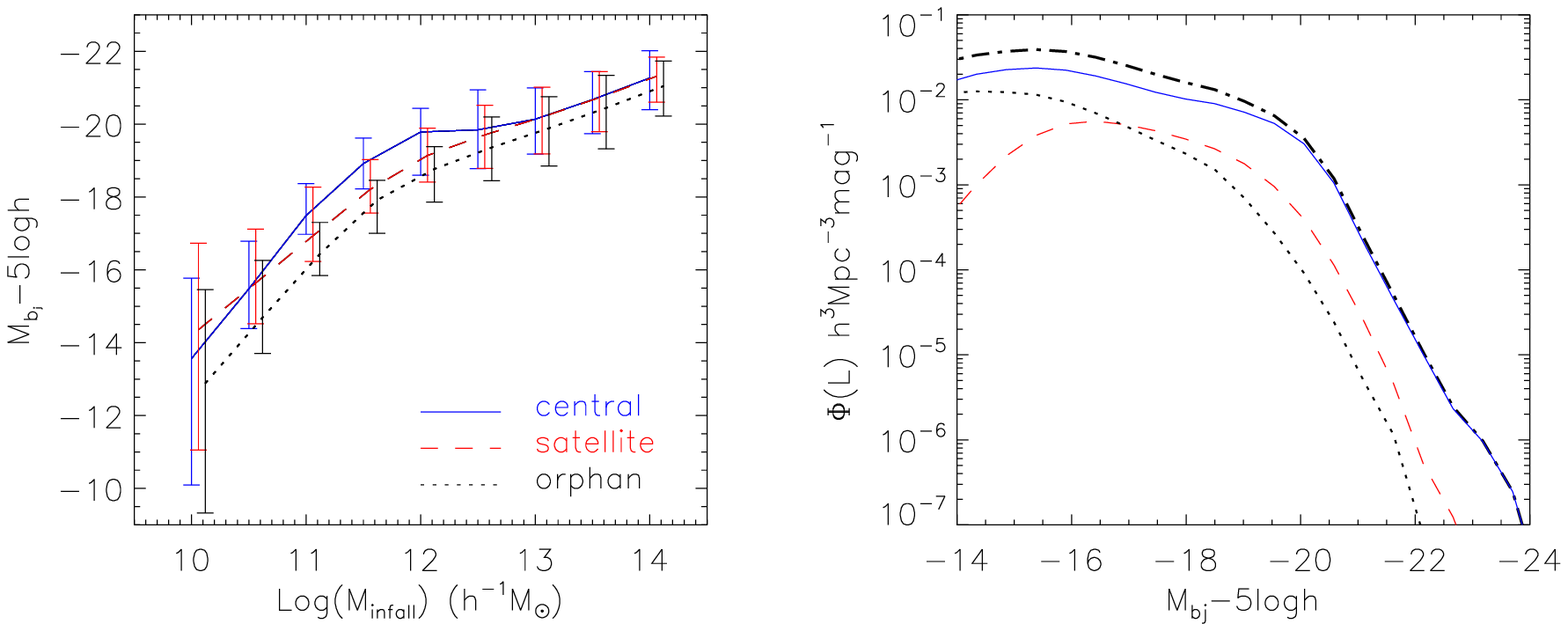}}\\%
\caption{$L-M_{infall}$ relations(left) and luminosity functions(right) for 
different 
types of galaxies from the semi-analytic galaxy catalogue: central galaxies
(solid lines), satellite galaxies with subhaloes (dashed 
lines), satellite galaxies without subhaloes (dotted lines). The
dashed-dotted line in the right-hand panel shows the total luminosity
function for all galaxies.}
\label{fig:type2}
\ec
\end{figure*}

\begin{figure*}
\bc
\hspace{-1.4cm}
\resizebox{17cm}{!}{\includegraphics{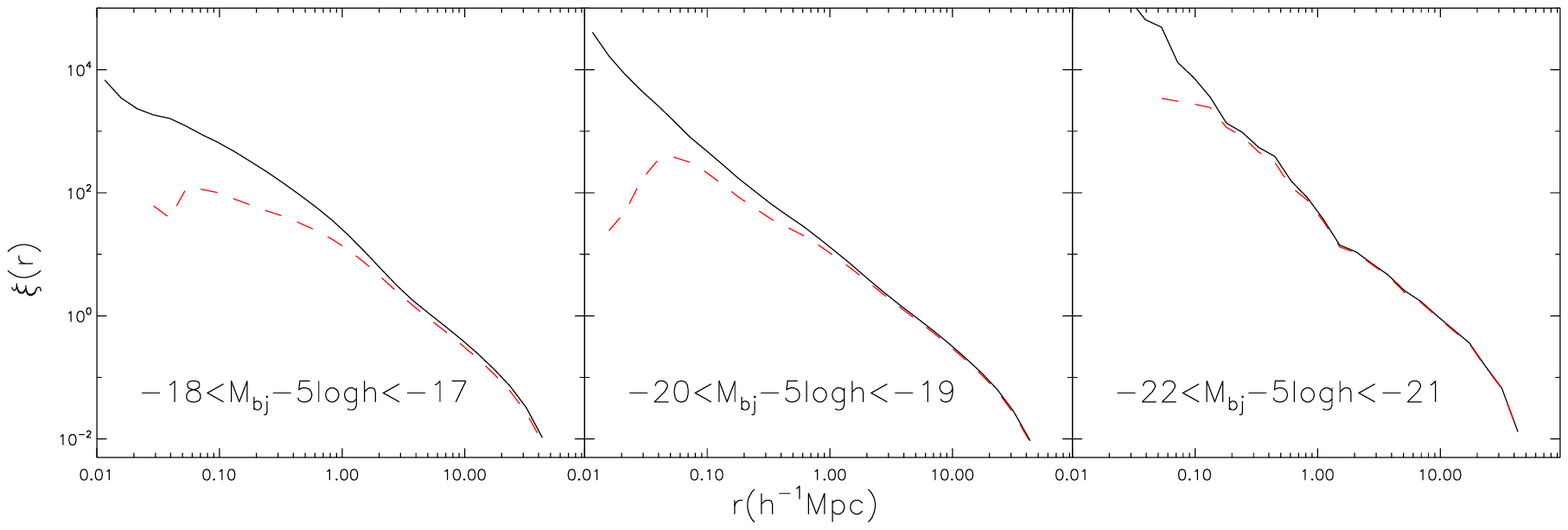}}\\%
\caption{Correlation functions for three luminosity bins   
including (solid lines) and not including (dashed lines) orphan satellite
galaxies.
}
\label{fig:type2corr}
\ec
\end{figure*}

\subsection {The effect of ``Orphan'' Galaxies}
The majority of HOD models in the literature only
consider dark matter haloes and subhaloes that can be identified at the
present time. Satellite galaxies without surrounding subhaloes are thus omitted
from the analysis. We now explore the effect of these 
'orphan' satellite galaxies on our results.

The left panel of Fig.~\ref{fig:type2} compares the 
$L-M_{infall}$ relation for orphan satellites with the results obtained for
central galaxies and satellite galaxies that have retained their
subhaloes. The right panel of Fig.~\ref{fig:type2} shows the relative  
contribution of central galaxies, satellite galaxies with subhaloes 
and orphan satellites without subhaloes to 
the luminosity function of the galaxies in the semi-analytic catalogue.
As can be seen, orphan satellites have lower luminosities at
a given value of $M_{infall}$ than either central galaxies or satellite
galaxies with subhaloes -- i.e. orphan galaxies are the oldest 
galaxies with the highest mass-to-light ratios  in the simulation. 
In addition, we see that the contribution of 
these orphan satellites is highest at the faintest lumninosities.
Fig.~\ref{fig:type2corr} explores the contribution of the orphan satellites
to the correlation function in three different bins of absolute
magnitude. The solid curves show the result for all the galaxies
while the dashed red curves show the result when the orphan satellites
are omitted. As can be seen, the orphaned satellites contribute
heavily to the correlation function of faint galaxies on scales
of less than 1 Mpc. Omission of these systems causes the amplitude
of the correlation function to be underestimated by more than an
order of magnitude at separations of 0.1 Mpc for galaxies with
$-18 < M_{bj} < -17$. 

We note that there are uncertainties in our treatment of
orphan galaxies in the simulation. Some of these galaxies may
indeed be destroyed or significantly reduced in mass by tidal
stripping effects. Indeed, the existence of a significant intra-cluster
light component does suggest tidal effects or mergers do unbind
some of the stars in satellite galaxies\citep{arnaboldi2004, 
feldmeier2004, zibetti2005}.
In face of these uncertainties, we have chosen to assume that the visible 
galaxies survive even after their subhalo falls below the resolution limit 
of the simulation.  It is possible that we over-estimate the number of these
objects because we do not include tidal stripping on the stellar component.
However, we believe that "orphan" galaxies (at least part of them) are needed
in order to explain observational results. From Fig.7 we see that when 
"orphan" galaxies are excluded, the correlation signal decreases at small 
scales, at odds with observational results(see later in Fig.9 and Fig.10).
In this work we consider all the 'orphan'
systems as part of satellite subsamples.  

\subsection {Changes in the input parameters}
One advantage of our parametrized approach is that we can understand the 
effect of changing each different parameter
and thus gain intuition about what changes are necessary
to bring the models into the closest possible agreement with
the observational data. This is different in spirit to exploring parameter
space in the semi-analytic models, because the parameters in these models
are tied to the physical recipes for star formation
and feedback rather than the relation between halo mass and
galaxy properties, which is the focus of our approach.

\begin{figure*}
\bc
\hspace{-1.4cm}
\resizebox{17cm}{!}{\includegraphics{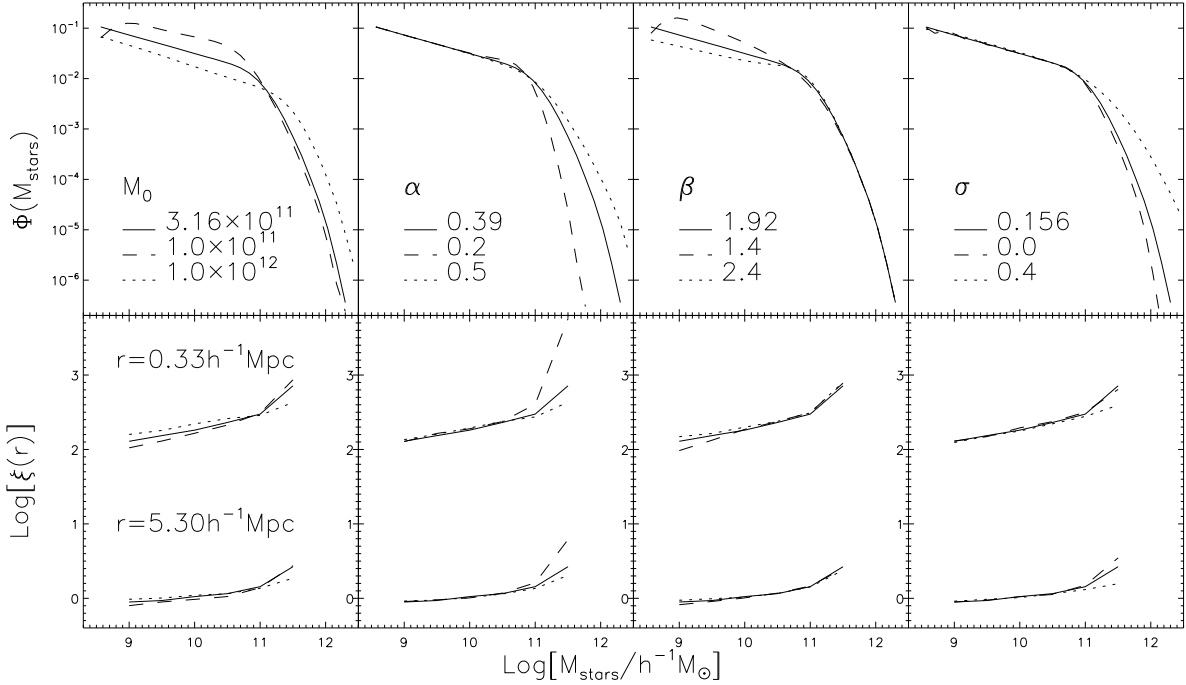}}\\%
\caption{
The effect of changing parameters on the stellar mass function(upper panels) 
and correlation at scales of $r=0.33h^{-1}Mpc$ and $r=5.30h^{-1}Mpc$(lower panels). The solid lines represent the best fit 
model for the $M_{stars}-M_{infall}$ relation.}
\label{fig:SMFchangepara}
\ec
\end{figure*}

In the upper panel of  Fig.~\ref{fig:SMFchangepara}, 
we show how changing each of the parameters
affects the stellar mass function. Note that the
normalization constant $k$ is always adjusted
in order to keep the amplitude of the mass function
at $M_{stars} = 10^{11} M_{\odot}$ fixed. Changing $M_0$
affects the mass scale of the transition between the
two power laws as well as the amplitude of the mass function
at both low and at high masses.  
Changes to $\alpha$ affect the shape of the mass function 
at the high mass end, 
while changes to $\beta$ affect the low mass end of the mass function.
A change in scatter $\sigma$ has similar effect to a change in
$\alpha$, and influences the amplitude of the mass function at
the high mass end. This is because the mass function is relatively
flat at low masses and  declines steeply at high masses, so
an increasing amount of scatter in the $ M_{stars}-M_{infall}$ relation
will have a strong effect on the number of high mass galaxies.

The lower panels in Fig.~\ref{fig:SMFchangepara} show the
effect of the same parameter changes on the amplitude of the correlation 
function evaluated on scales of
$r=0.33h^{-1}$ Mpc and $r=5.30h^{-1}$ Mpc.
We see that a parameter change that causes an increase in the number
of galaxies of given mass, will cause a corresponding {\em decrease}
in the clustering amplitude of these systems. This is easy to understand. 
In order to have more galaxies of a given mass in the simulation,
they must be shifted into lower mass haloes and these low mass
haloes are more weakly clustered.

% %%%%%%%%%%%%%%%%%%%%%%%%%%%%%%%%%%%%%%%%%%%%%%%%%%%%%%%%%%%%%%%%%%
\section{Application to SDSS}
\label{sec:sdss}

In this section, we apply our models to observational data from the Sloan
Digital Sky Survey. Recent
large scale redshift surveys such as 2dfGRS\citep{colless01} and Sloan 
Digital Sky Survey(SDSS; \citet{york2000}) provide galaxy samples that
are large enough to measure the luminosity dependence of galaxy clustering 
accurately \citep{norberg02a,norberg02b,zehavi2005}. 
In this paper, we make use of the recent measurements of the
projected correlation function $w(r_p)$ by \cite{li2005a}. These authors
calculated $w(r_p)$ not only as a function of galaxy luminosity, but
also stellar mass using a sample of galaxies constructed from the
SDSS Data Release 2 (DR2) data. The methods for estimating the stellar
masses are described in \cite{kauffmann2003}.
Here we make use of these measurements to constrain the 
relation between galaxy luminosity, stellar mass and $M_{infall}$.
To take account the effect of "cosmic variance" on  the observational
results, we have constructed a set of 16 mock galaxy catalogues from the 
simulation with exactly the same geometry and selection function as in the 
observational sample. The effect of  cosmic variance is modelled by 
placing a virtual observer randomly inside the simulation box when 
constructing these mock catalogues. For each mock catalogue, we measure 
$w(r_p)$ for galaxies in the same intervals of luminosity/stellar mass as 
in the observations. 
The $1-\sigma$ variation between these mock catalogues is then added as an 
additional error in quadrature to the bootstrap errors given by 
\citet{li2005a}. The cosmic variance errors become significant for the
low luminosity and low mass subsamples, particularly at large values
of $r_p$. The detailed procedure for constructing these mock catalogues 
will be presented in a separate paper (Li et al., in preparation).

\begin{figure*}
\bc
\hspace{-1.4cm}
\resizebox{16cm}{!}{\includegraphics{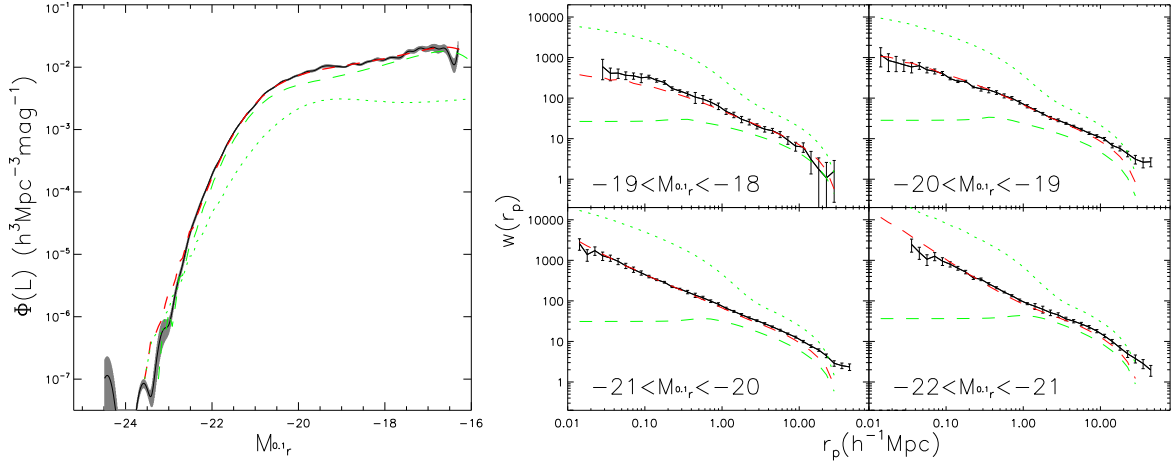}}\\%
\caption{Best fit model to the  luminosity function and the correlation 
function evaluated in different luminosity bins using data from
the SDSS. Solid lines with error bars are the SDSS results, and red dashed 
lines are from our parametrized model. Green dashed/dotted lines are results 
for central/non-central subsamples of our parametrized model.
   }
\label{fig:SDSSlf}
\ec
\end{figure*}

\begin{figure*}
\bc
\hspace{-1.4cm}
\resizebox{16cm}{!}{\includegraphics{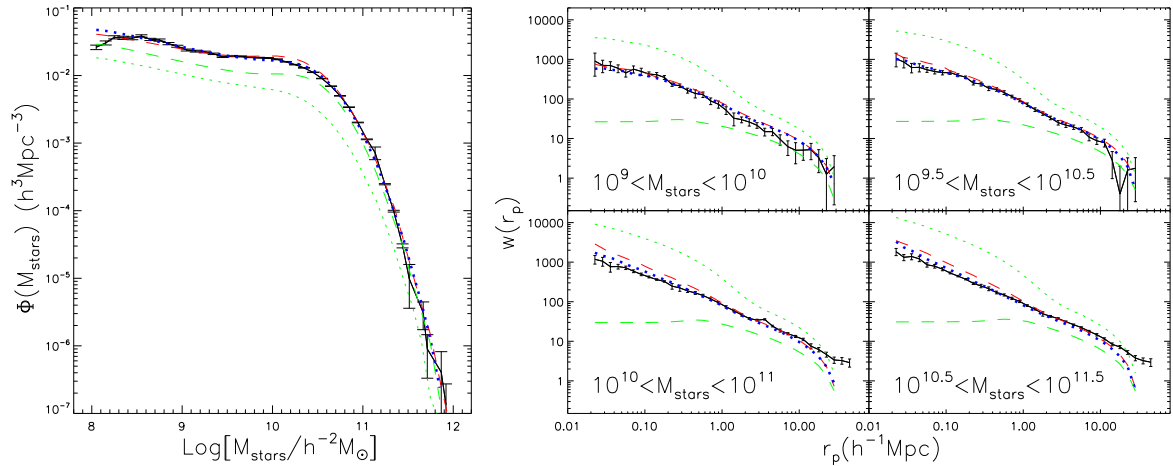}}\\%
\caption{Best fit model to the stellar mass function and the correlation 
function evaluated in different stellar mass bins using
data from the SDSS. Symbols with error bars are the SDSS results, 
and dashed red lines are from our parametrized model.
Dotted blue lines show the results obtained when central and 
satellite galaxies are treated separately. Green dashed/dotted lines 
are results for central/non-central subsamples of the parametrized model 
when central and satellite galaxies are treated separately.}
\label{fig:SDSSsmf}
\ec
\end{figure*}

To compare our models with the observations, we 
need to either convert $w(r_{p})$ to the real space correlation function
$\xi(r)$, or to calculate $w(r_{p})$ from our model galaxy catalogue
directly. We tested the method presented by \citet{hawkins2003} for converting 
$w(r_{p})$ to $\xi(r)$ on scales less than 
around $30h^{-1}Mpc$. We find that the conversion amplifies the error
and the results for the low luminosity and low mass bins are then too noisy
to provide good constraints on our models. 
Therefore, we derive $w(r_{p})$ from our catalogue
by integrating the real space correlation function $\xi(r)$:
 \begin{displaymath}
w(r_{p})={2}{\int_{0}^{\infty}{\xi(\sqrt{{r_{p}}^{2}}+{r_{\parallel}}^{2}){d}{r_{\parallel}}}}={2}{\int_{r_{p}}^{\infty}{\xi(r)}\frac{rdr}{\sqrt{r^{2}-{{r_{p}}^{2}}}}}
\end{displaymath}
We truncate the integration at $r=60h^{-1}Mpc$  and the resulting $w(r_{p})$  
is reliable up to a scale of $\sim10h^{-1}Mpc$.

\begin{figure*}
\bc
\hspace{-1.4cm}
\resizebox{16cm}{!}{\includegraphics{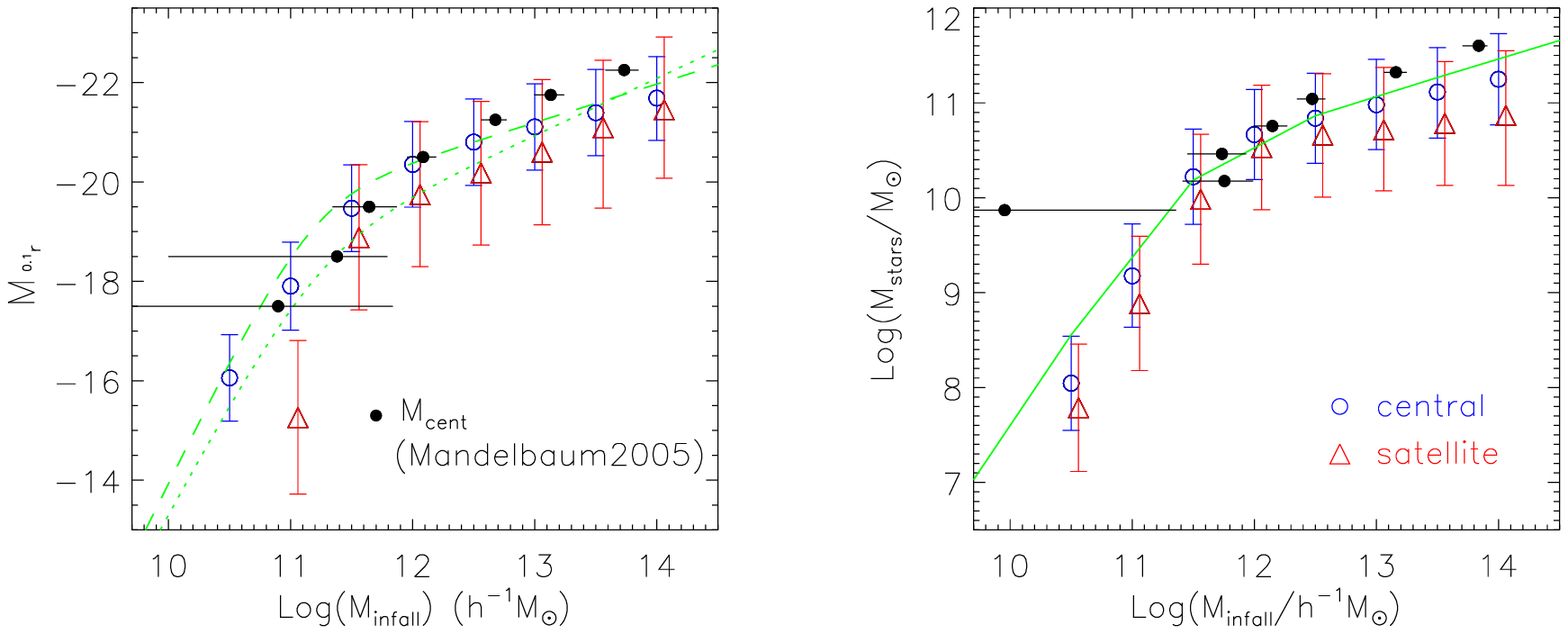}}\\%
\caption{Best fit $L-M_{infall}$ and $M_{stars}-M_{infall}$ relations 
as constrained by the SDSS data. Blue circles are
the central galaxies and red triangles are satellites. Green lines are 
the best fitting relations from the semi-analytic catalogue of 
\citet{croton2005}. Filled circles show the central halo mass from the 
galaxy-galaxy lensing results of \citet{mandelbaum05b}; error
bars are the $95\%$ confidence limits. The results shown are the combined 
sample of early and late-type galaxies(Mandelbaum, private communication).}
\label{fig:SDSSMMinfall}
\ec
\end{figure*}

We now generate a grid of models by systematically varying the 5
parameters listed in Table 1. We compare each model with the
galaxy luminosity function\citep{blanton2003b} and the $w(r_p)$
measurements in different ranges in luminosity.
We define the best fitting model to be the one giving a minimum 
$\chi^2$ defined as follows:                                               
\begin{displaymath}
{{\chi}^2}= \frac{{\chi}^2(\Phi)}{N_{\Phi}}+\frac{{\chi}^2_{corr}}{N_{corr}}
\end{displaymath}
with
\begin{displaymath}
{{\chi}^2(\Phi)}= \sum_{N_{\Phi}}{[\frac{\Phi-\Phi_{SDSS}}{\sigma(\Phi_{SDSS})}]^2}
\end{displaymath}
and
\begin{displaymath}
{{\chi}^2_{corr}}= \sum_{N_{corr}}{[\frac{w(r_p)-w(r_p)_{SDSS}}{\sigma(w(r_p)_{SDSS})}]^2}
\end{displaymath}

$N_{\Phi}$ is the number of points over which the luminosity
function is measured 
($N_{\Phi}=102$ for the r-band absolute magnitude ranging from  
$-18$ to $-23$). $N_{corr}$ is the number of points
over which the correlation function is measured ( $N_{corr}=93$, 
ranging from $0.11$ to 
$8.97h^{-1}$ Mpc for luminosity bins$[-19,-18], [-20,-19], [-21,-20], 
[-22,-21]$ and from $0.57$ to $8.97h^{-1}$ Mpc for the most luminous bin
$[-23,-22]$ in the  $r$-band). 

To compare with the SDSS observations, where the median
galaxy redshift is around $0.1$, 
we correct the $r$-band absolute magnitude $M_r$ of each model galaxy
to its $z=0.1$ value $M_{^{0.1}r}$ using the $K-$correction code 
({\tt kcorrect v3\_1b}) of \citet{blanton2003a} and the luminosity 
evolution model of Blanton et al. (2003b). To calculate the K-
and E-correction, each galaxy is assigned a redshift by
placing a virtual observer at the centre of the simulation box.
The redshift as "seen" by the observer is thus determined by the comoving
distance to the observer and the peculiar velocity of the galaxy.
The corrected $r$- band magnitude  
is given  by:
\begin{displaymath}
{M_{^{0.1}r}}= {{-2.5}{\times}{\log}{L}+{K_{correction}}+{E_{correction}}-{5}{\log}{h}}
\end{displaymath}

 \begin{table*}
 \caption{Best-fit parameter values for the relations between
$M_{infall}$ and $M_{stars}$ and $L (M_r)$ as derived
from the SDSS data.}
\begin{center}
 \begin{tabular}{cccccccccc} \hline
   &                      &  $M_0(h^{-1}M_{\odot})$ & $\alpha$ & $\beta$ & $log(k)$ & $\sigma$& $\chi^2$ & $\chi^2(\Phi)/N_{\Phi}$\\ \hline
 $M_{stars}(M_{\odot})$ & total     & 3.15$\times10^{11}$ & 0.118 & 2.87 & 10.26    & 0.326   & 16.96    & 2.487\\
                        & central   & 3.33$\times10^{11}$ & 0.276 & 2.59 & 10.27    & 0.241   & 5.351    & 1.850\\
                        & satellite & 4.64$\times10^{11}$ & 0.122 & 2.48 & 10.26    & 0.334   & &\\ \hline
 $L(M_r)$               & central   & 3.41$\times10^{11}$ & 0.221 & 1.67 & 8.13     & 0.440   & 6.115    & 3.348\\
                        & satellite & 2.58$\times10^{11}$ & 0.345 & 3.83 & 7.71     & 0.742   & &\\ \hline
 \end{tabular}
\end{center}
 \end{table*}

Our best fit model has the parameters: $M_0=3.41\times10^{11}h^{-1}M_{\odot}$, 
$\alpha=0.221$, $\beta=1.67$, $k=8.13$ and $\sigma=0.440$ for the central 
galaxies and $M_0=2.58\times10^{11}h^{-1}M_{\odot}$, $\alpha=0.345$, 
$\beta=3.83$, $k=7.71$ and $\sigma=0.742$ for the satellite galaxies (see 
Table 2). The resulting luminosity function and correlation functions are  
shown in Fig.~\ref{fig:SDSSlf}. $\chi^2(\Phi)/N_{\Phi}$ is 
$3.348$ and the total $\chi^2$ is $6.115$. Also plotted are the results
of central and satellite subsamples of our parametrized model, 
shown by green dashed and dotted lines.
The drop in the correlation function on scales larger than $\sim10h^{-1}$ Mpc
is not caused by a poor fit; it is due to the truncation of our integration 
of the real space correlation function at $r= 60 h^{-1}$ Mpc$^{-1}$.

We now carry out the same analysis for stellar mass, rather than luminosity.
We have constructed the stellar mass function directly from the SDSS DR2 
data (Fig.~\ref{fig:SDSSsmf}; left panel) and use this, in conjunction
with the measurements of $w(r_p)$ as a function of stellar mass published by
\citep{li2005a}, to constrain the $M_{stars}$-$M_{infall}$ relation.
In the computation of stellar mass function, we have corrected for the 
volume effect by weighting each galaxy by a factor of $V_{survey}/V_{max}$, 
where $V_{survey}$ is the volume for the sample and $V_{max}$ is the maximum
volume over which the galaxy could be observed within the sample redshift
range ($0.01<z<0.3$) and within the range of $r-$band apparent magnitude
($14.5<r<17.77$). 
A Schechter function provides a good fit to our measurement
at stellar mass $M_{stars}<10^{11.5}h^{-2}M_\odot$. We find best-fit 
parameters: $\Phi^\ast=(0.0204\pm0.0001) h^3$Mpc$^{-3}$,
$\alpha=-1.073\pm0.003$ and $M^\ast_{stars}=(4.11\pm0.02)\times10^{10} h^{-2}M_\odot$. This corresponds to a stellar mass density of 
$(8.779\pm 0.067)\times 10^8 hM_{\odot}$Mpc$^{-3}$.

We fit our models to $30$ points along the stellar mass function
 and $20$ points along the correlation function for five different
stellar mass bins ranging from $10^9$ to $10^{12}M_{\odot}$. 
The parameters of the best-fit models are listed in Table 2.
For the stellar mass function, the errors due to sample size are much smaller
than the systematic errors in the stellar mass estimates themselves.
We therefore assign the same error to all points at stellar masses
less than 
$10^{11.5}h^{-2}M_{\odot}$ (the error is equal to the value at that mass). 
In our first attempt at fitting the data, we assumed that the
$M_{stars}-M_{infall}$ would be the same for central and 
satellite galaxies, because the relations are very similar in the semi-analytic
galaxy catalogues. The red dashed lines in Fig.~\ref{fig:SDSSsmf} show 
the best fitting results. The model clearly over-predicts the clustering
of the more massive galaxies on small scales. If we allow the relation
between $M_{stars}$ and $M_{infall}$ to differ for central
and satellite galaxies, we obtain the results shown by the blue dotted
lines, which are considerably better.     

The best-fit $r$-band luminosity -- $M_{infall}$ and 
$M_{stars}-M_{infall}$ relations  derived from our models are illustrated   
in Fig.~\ref{fig:SDSSMMinfall}. Results are shown separately
for central galaxies (blue) and satellite galaxies (red). In our  
models, satellite galaxies
have lower luminosities and smaller stellar masses than 
central galaxies at a given value
of $M_{infall}$. This effect is larger for luminosity than
for stellar mass, particularly at low values of $M_{infall}$.

For comparison, we also plot the $L$-$M_{infall}$ and  $M_{stars}-M_{infall}$  
relations from the semi-analytic galaxy catalogue\citep{croton2005}.
We transform the $b_j$ band magnitude of semi-analytic catalogue to $r$ band 
in SDSS according to the luminosity functions of 2dFGRS\citep{madgwick2002} 
and SDSS\citep{blanton2003b}, and make a shift of $0.9$ dex to do the 
comparison. Because Croton et al assumed a Salpeter initial mass function, the
stellar mass-to-light ratios of the galaxies in their catalogue
will be a factor of $\sim 2$ higher than in the SDSS data sample. 
This is because the stellar masses of \cite{kauffmann2003} have been derived 
using  a \citet{kroupa2001} IMF. As discussed by \citet{kauffmann2003}, the 
Salpeter IMF yields stellar masses for elliptical galaxies  that {\em exceed} 
estimates of their dynamical masses\citep{cappellari2006}. 
The Salpeter IMF is clearly unphysical and should be dropped. For the 
comparison shown in Fig.~\ref{fig:SDSSMMinfall}, we have simply
scaled the stellar masses in the Croton et al catalogues by multiplying
by a factor 0.5, which should give almost the same results as re-running
the semi-analytic model with the Kroupa IMF.
Compared with our results, the semi-analytic catalogue yields
systematically higher luminosities and  stellar masses 
at low values of  $M_{infall}$, particularly for satellite galaxies.
The agreement with the semi-analytic catalogue at $M_{infall} >
10 ^{11.5} h^{-1} M_{\odot}$ is quite good.

Recently \citet{mandelbaum05} and \citet{mandelbaum05b}  
have used galaxy-galaxy weak lensing
measurements from SDSS data to explore the 
explore the connection between galaxies and dark matter. 
They compare the predicted lensing signal 
from a halo model constructed using a dissipationless simulation, and 
extract median/mean halo masses and satellite fractions for
galaxies as a function of luminosity, stellar mass and morphology.
We plot their estimates of the mean central halo mass as a function
of $r$-band absolute magnitude and stellar mass as filled circles in
 Fig.~\ref{fig:SDSSMMinfall}.  The results shown are the combined 
sample of early and late-type galaxies(Mandelbaum, private communication).
These measurements should be
compared with our blue points, which show the mean halo masses
of present-day central galaxies. As can be seen,
there is remarkably good agreement between the two methods, both 
for luminosity and for stellar mass.

% %%%%%%%%%%%%%%%%%%%%%%%%%%%%%%%%%%%%%%%%%%%%%%%%%%%%%%%%%%%%%%%%%%
\section{ Conclusions and Discussion }
\label{sec:conclusion}

We have constructed a new statistical model of galaxy clustering
for use in high resolution numerical simulations of structure formation.
Unlike classic halo occupation distribution (HOD) models, 
galaxy positions and velocities are determined in a self-consistent way
by following the full orbital and  merging histories of all the haloes
and subhaloes in the simulation. We believe that this methodology has 
advantages over the traditional approach. Most HOD models assume that the   
galaxy content of a halo of given mass is statistically 
independent of its larger scale environment. Recently \citep{gao2005} have  
shown that there exists an age dependence of halo 
clustering: haloes that are formed earlier are more clustered than haloes 
that are assembled more recently, indicating that this assumption may not be
as safe as previously thought. Since the positions and the velocities of the 
galaxies in our model are determined directly from the simulation, we avoid 
these difficulties.

Our methodology also takes into account the contribution
of ``orphaned'' galaxies, which have lost their halos due to tidal
stripping. These galaxies contribute significantly to
the clustering amplitude of low mass galaxies on scales less than 
$\sim 1 h^{-1}$ Mpc.
We have chosen to parametrize the observed properties of galaxies  
(in particular their luminosity and stellar mass) 
as a function of the quantity $M_{infall}$, the mass of the halo at the
epoch when the galaxy was last the central object in its halo. 
Using the semi-analytic model results as a reference, we adopt a double 
power law   
form for this relation, and we show that this allows us to recover 
the mass/luminosity function and the correlation function in different
ranges of mass and luminosity with high accuracy.  

We then apply our model to measurements of these quantities using data
from the Sloan Digital Sky Survey. We find that for a given value of
$M_{infall}$, satellite galaxies are required to be less luminous and less 
massive than central galaxies. This effect is stronger at low values of
$M_{infall}$. In the semi-analytic models, satellite galaxies 
fade in luminosity after they fall into a larger halo because they no longer
accrete gas and their star formation rates then decline. The catalogues
of \citet{croton2005} do show differences between satellite and
central galaxy luminosities at a fixed value of $M_{infall}$, but
the effect is not quite as strong as the data demands,
particularly for low mass halos. This may indicate that
the efficiency with which baryons are converted into stars
in low mass halos is higher at the present day than it was in the past.
The fact that the standard $\Lambda$CDM model predicts more 
low mass galaxies than observed is very well-documented in the
literature \citep{moore1999}. 
Many authors have tried to invoke mechanisms for ``suppressing''
star formation in these systems \citep{kauffmann1993,somerville2002}
and most of these mechanisms operate more effectively at higher redshifts.

Finally, we compare our relations between galaxy luminosity, stellar mass and
host halo mass with similar relations derived using galaxy-galaxy
weak lensing measurements. The excellent agreement between these
two completely independent methods is very encouraging.

%%%%%%%%%%%%%%%%%%%%%%%%%%%%%%%%%%%%%%%%%%%%%%%%%%%%%%%%%%%%%%%%%%%
\section*{Acknowledgements}

We are grateful to Simon White, Darren Croton and the referee 
J. S Bullock for their detailed comments and suggestions on our paper. 
We thank Rachel Mandelbaum for 
providing their results on Fig 11. C.~L acknowledges the financial 
support of the exchange program between Chinese Academy of Sciences
and the Max Planck Society. G.~D.~L. thanks the Alexander von Humboldt 
Foundation, the Federal Ministry of Education and Research, and the 
Programme for Investment in the Future (ZIP) of
the German Government for financial support.

The simulation used in this paper was carried out as part of the programme of 
the Virgo Consortium on the Regatta supercomputer of the Computing Centre of 
the Max--Planck--Society in Garching.

Funding for the SDSS and SDSS-II has been provided by the Alfred P. Sloan Foundation, 
the Participating Institutions, the National Science Foundation, 
the U.S. Department of Energy, the National Aeronautics and Space Administration, 
the Japanese Monbukagakusho, the Max Planck Society, and the 
Higher Education Funding Council for England. The SDSS Web Site is http://www.sdss.org/. The SDSS is managed by the Astrophysical Research Consortium 
for the Participating Institutions. The Participating Institutions are 
the American Museum of Natural History, Astrophysical Institute Potsdam, 
University of Basel, Cambridge University, 
Case Western Reserve University, University of Chicago, 
Drexel University, Fermilab, the Institute for Advanced Study, 
the Japan Participation Group, Johns Hopkins University, 
the Joint Institute for Nuclear Astrophysics, the 
Kavli Institute for Particle Astrophysics and Cosmology, the 
Korean Scientist Group, the Chinese Academy of Sciences (LAMOST), 
Los Alamos National Laboratory, 
the Max-Planck-Institute for Astronomy (MPIA), 
the Max-Planck-Institute for Astrophysics (MPA), 
New Mexico State University, Ohio State University, 
University of Pittsburgh, University of Portsmouth, 
Princeton University, the United States Naval Observatory, 
and the University of Washington.

\bsp
\label{lastpage}

\bibliographystyle{mn2e}
\bibliography{HOD_wanglan}

\end{document}